\documentclass[twocolumn]{autart}    

\usepackage{cite}
\usepackage{amsmath,amssymb,amsfonts}
\usepackage{algorithmic}
\usepackage{graphicx}
\usepackage{textcomp}

\usepackage{graphicx}          
\usepackage{epstopdf}
\begin{document}

\begin{frontmatter}

\title{Arbitrary State Transition of Open Qubit System Based on Switching Control} 

\thanks[footnoteinfo]{This work is supported by the National Natural Science Foundation of China (NSFC) under Grants 62273226.}

\author[GG1,GG2,GG3]{Guangpu Wu},
\author[GG1,GG2,GG3]{ Shibei Xue},\ead{shbxue@sjtu.edu.cn}              %
\author[GG4]{ Shan Ma},
\author[GG5]{ Sen Kuang},
\author[GG6]{ Daoyi Dong},
\author[GG6]{ Ian R. Petersen}

\address[GG1]{Department of Automation, Shanghai Jiao Tong University, Shanghai 200240, P. R. China}
\address[GG2]{Key Laboratory of System Control and Information Processing, Ministry of Education of China, Shanghai 200240, P. R. China}
\address[GG3]{Shanghai Engineering Research Center of Intelligent Control and Management, Shanghai 200240, P. R. China}
\address[GG4]{School of Automation, Central South University, Changsha 410083,  P. R. China}
\address[GG5]{Department of Automation, University of Science and Technology of China, Hefei 230027, PR China}
\address[GG6]{School of Engineering,  Australian National University, Canberra, ACT 2601, Australia}
%
%


\begin{abstract}
We present a switching control strategy based on Lyapunov control for arbitrary state transitions in open qubit systems.
With coherent vector representation, we propose a switching control strategy, which can prevent the state of the qubit from entering invariant sets and singular value sets, effectively driving the system ultimately to a sufficiently small neighborhood of target states.
In comparison to existing works, this control strategy relaxes the strict constraints on system models imposed by special target states.
 Furthermore, we identify conditions under which the open qubit system achieves finite-time stability (FTS) and finite-time contractive stability (FTCS), respectively.
 This represents a critical improvement in quantum state transitions, especially considering the asymptotic stability of arbitrary target states is unattainable in open quantum systems.
 The effectiveness of our proposed method is convincingly demonstrated through its application in a qubit system affected by various types of decoherence, including amplitude, dephasing and polarization decoherence.
%
\end{abstract}

\end{frontmatter}

\section{Introduction}

\par{
With the help of principles of quantum mechanics, quantum information processing, including quantum communication \cite{5.01} and quantum control \cite{5.1,5.2}, surpasses classical information technics in some aspects \cite{6,7}.
 As the basic carrier of quantum information, qubit systems play a crucial role in quantum information processing \cite{7,8}. Hence, precise control of qubit states is not only fundamental to the success of quantum communication and control but also pivotal in advancing the field of quantum information technologies \cite{5.3,5.4}. A fundamental control problem is: how to guide the state of qubit systems toward a target state via suitable controls \cite{11,12}.
 }

\par{
To tackle the problem, a class of Lyapunov control methods has been proposed, where Lyapunov stability is utilized to design control laws, stabilizing a system to a target state \cite{13,14}. The challenge of quantum Lyapunov theory lies in the design of a Lyapunov function \cite{16,17}.
For closed systems, Ref.\cite{15.1} presents three distinct Lyapunov functions based on state distance, average value of an
imaginary mechanical quantity and state error, which are commonly used in existing works. Based on the Lyapunov function of state distance method in \cite{15.1}, Ref.\cite{18} has proposed target state tracking control for closed systems, which can be considered as a state transition problem when the target state is stationary. However, it only works for  a specific type of target state which is a diagonal density matrix. Extending the results in \cite{18}, Ref.
\cite{19.1} takes the eigenstates of the free non-degenerate Hamiltonian as the target states.
Furthermore, to transfer to eigenstates of the Hamiltonian, advanced control methods are explored to accelerate the speed of state transitions \cite{19.11}, achieving state transfers within a finite time through non-smooth control laws \cite{19.2}.
}

\par{
Different from closed quantum systems, controlling open quantum systems is usually more challenging \cite{17.0,17.00}.
 To achieve global asymptotic stability in state transition, target states have to be chosen as special states.
%
For open quantum systems under measurement feedback described by a quantum stochastic master equation, Ref.\cite{17.1} adopted the eigenstates of the Hamiltonian as target states and presents a segmented control approach, where different control laws are designed based on the proximity to the target states, ultimately stabilizing the system in a probabilistic sense.
 Similar to \cite{17.1},
 Ref.\cite{17.4} reconfigured the control Hamiltonian, using segmented control to avoid the state entering into sets of singular values and invariant sets, making it possible to prepare entangled states such as GHZ and W states in qubit systems.
Similarly, Ref.\cite{17.5} investigates state transition under time-delay conditions.
In all the aforementioned works with measurement feedback, the target states are eigenstates of an observable, and segmented control is employed to ensure the stability in probability. For systems without measurement feedback, Ref.\cite{17.51} investigates a switching approach to driving a state into a decoherence-free space (DFS).
%
However, to achieve an asymptotically stable convergence, most existing works focus on specific target states. Here, we focus on the state transition problem for arbitrary target states where the existing control laws often do not work.
}

\par{
In this paper, we investigate the arbitrary state transition problem in open qubit systems where we do not require the target state to be the eigenstate of the system Hamiltonian or commute with Lindblad operators. A switching control strategy is employed to prevent the state entering invariant and singular value sets.
 However, given that open qubit systems cannot achieve asymptotic stability for an arbitrary target state \cite{29}, we propose a strategy to direct the state into a small neighborhood encompassing the target state through introducing the concepts of finite-time stability (FTS)
and finite-time contraction stability (FTCS). These concepts consider practical constraints and the need for a swift response, aiming to ensure that the system attains and
maintains proximity to the target state within a finite
time.
Building upon this, we prove the convergence of the system state when employing the proposed control strategy.
Finally, the proposed method is applied to three distinct decoherence systems, validating its effectiveness.
}

This paper is organized as follows. Section 2 introduces a qubit control model, presents the Bloch vector representation and formulates the arbitrary state transition problem. Section 3 introduces our Lyapunov control method and switching control strategy. In Section 4, the finite time convergence analysis for the problem is given. Section 5 demonstrates the efficacy of the switching control strategy through simulations in various decoherence cases. Finally, conclusions are drawn in Section 6.

\textbf{Notation} For a matrix $A=[A_{ij}]$, the symbols $A^T$ and $A^\dag$ represent the transpose and Hermitian conjugate of $A$. Given two operators $M$ and $N$, $[M,N]=MN-NM$ and $\{M,N\}=MN+NM$ are their commutators and anti-commutators, respectively. Given a complex number $a$, $\bar a$ represents its conjugate.
\section{State transition for open qubit systems}
\subsection{Open single qubit systems}
The state of a single qubit system defined on a two dimensional Hilbert space $\mathcal{H}$ can be represented by a density operator $\rho\in S=\{\rho|\rho^\dag=\rho,\rho\geq0,\mathrm{tr}(\rho)=1,\rho\in\mathbb{C}^{2\times2}\}$. The evolution of a Markovian open qubit system obeys the following Markovian master equation \cite{17.6,17.7}
\begin{align}\label{drho}
\dot \rho(t)=-\frac{i}{\hbar}[H,\rho(t)]+\sum_{j=1}^R\gamma_j\mathcal{D}_{L_j}[\rho(t)],
\end{align}
where $H=H_0+\sum_{r=1}^2u_r(t)H_r$ with free Hamiltonian $H_0$ and control Hamiltonians $H_r$. $u_r(t)$ are admissible real-valued control functions. The reduced Planck constant $\hbar$ is set to be $1$. Generally, the first term on the RHS of Eq.\eqref{drho} describes the unitary evolution of the qubit without interaction with an environment. The positive time-independent parameters $\gamma_j\in\mathbb{R}^+$ represent the dissipation rates in different damping channels \cite{20}. $R$ represents the number of damping channels. The Lindblad superoperator
\begin{align}
\mathcal{D}_{L_j}[\rho(t)]=L_j\rho(t) L_j^\dag-\frac{1}{2}(L_j^\dag L_j\rho(t)+\rho(t) L_j^\dag L_j)
 \end{align}
 is induced by a Markovian environment,
where $L_j$ are Lindblad operators.
\subsection{Bloch vector representation}
To obtain a dynamical equation with real-valued coefficient matrices, the density matrix describing quantum states can be converted into a coherent vector representation \cite{26,27}.
  In this representation, the state of the system \eqref{drho} can be expanded as
     \begin{equation}
\rho(t)=\frac{1}{2}(I_2+\sum^3_{v=1}s_v(t)\sigma_v),
       \end{equation}
       where $s_v(t)=\mathrm{tr}(\rho(t)\sigma_v)\in\mathbb{R}$, $I_2$ is a two dimensional identity matrix. The Pauli matrices
\begin{align}\label{H}
\sigma_1=\left[\begin{array}{cc}0&1  \\1 &0\\\end{array}\right],
\sigma_2=\left[\begin{array}{cc}0&-i  \\i &0\\\end{array}\right],
\sigma_3=\left[\begin{array}{cc}1&0  \\0 &-1\\\end{array}\right]
\end{align}
are the orthonormal bases for the Lie algebra $\mathfrak{su}(2)$. Their commutation relations are given as
       \begin{align}\label{re1}
[\sigma_m,\sigma_n]=2i\epsilon_{mnl}\sigma_l,
\end{align}
where $m,n,l\in\{1,2,3\}$, the anticommutation relations are
       \begin{align}\label{re2}
\{\sigma_m,\sigma_n\}=2\delta_{mn}.
\end{align}
where $\epsilon_{mnl}$ is the completely antisymmetric structure constant of the Lie algebra $\mathfrak{su}(2)$ and $\delta_{mn}$ represents Kronecker delta function.  Similarly, the Hamiltonians and Lindblad operators can be expanded as
            \begin{equation}\label{hq}
H_q=\sum^3_{v=1}h^q_v\sigma_v,
       \end{equation}
                \begin{equation}\label{lj}
L_j=\sum^3_{v=1}\beta^j_v\sigma_v,
       \end{equation}
       where $h^q_v=\mathrm{tr}(H_q\sigma_v)$, $\beta^j_v=\mathrm{tr}(L_j\sigma_v)$, $q=0,1,2$.
      Hence, the Lindblad master equation \eqref{drho} can be reformulated into a Bloch equation
 \begin{equation}\label{vect}
\dot s(t)=As(t)+\sum_{r=1}^2 u_rK_rs(t)+g,
       \end{equation}
       where $s(t)=[s_1(t),s_2(t),s_3(t)]^T$ is the coherent vector; $A$ and $K_n\in \mathbb{R}^{3\times 3}$ are the system and control matrices, respectively; $g=[g_1,g_2,g_3]^T\in \mathbb R^3$ is a time-independent constant vector. The elements of $A$, $K_r$ and $g$ can be calculated as
\begin{align}
&A_{ll}=-2\sum_{j=1}^R\sum^3_{m\not=l}\gamma_j|\beta^j_m|^2 ,\label{vec1}\\
&A_{lm}=2\sum_{p=1}^3 h^0_p\epsilon_{pml}+\sum_{j=1}^R\gamma_j (\beta^j_l\bar{\beta^j_m}+\beta^j_m\bar{\beta^j_l}),\label{vec2}\\
&K^r_{ll}=0,\\
&K^r_{lm}=2\sum_{p=1}^3 h^r_p\epsilon_{pml},\\
&g_{l}=2i\sum_{j=1}^R\sum_{m=1}^3 \sum_{p=1}^3 \gamma_j\beta^j_m\bar{\beta^j_p}\epsilon_{mpl}.\label{vec5}
\end{align}
In fact, the evolution of the state vector of the system \eqref{vect} and the density matrix of system \eqref{drho} has an one-to-one mapping. In the following, we frequently treat the two equations as equivalent.

      %


\subsection{Problem formulation}
Different from Refs. \cite{17.1,17.4,17.5}, the target state $\rho_d$ in this paper is not an eigenstate of the Hamiltonian or observables, nor is it within a DFS. Instead, it is relaxed to be an arbitrary quantum state.
 Since the system \eqref{vect} are unable to attain asymptotic stability for an arbitrary target state \cite{29}, our aim is to stabilize the system state within a sufficiently small neighborhood through control within a finite time.
Based on Eq. \eqref{vect}, the initial state $\rho_0$ and target state $\rho_d$ can be thus denoted as $s_0$ and $s_d$ under the representation of the coherent vector. The problem of this paper is to design control laws $u_1(t)$ and $u_2(t)$ such that the initial state $s_0$ converges to a neighborhood of an arbitrary target state $s_d$ within a finite time.



\section{Lyapunov control}
In order to transfer the state of the system \eqref{vect} to a target state, we provide a Lyapunov control method based on state error in this section. With this method, a switching control strategy is proposed.
\subsection{Lyapunov control design for arbitrary state transition}
Firstly, we define an error vector $e(t)$ between the state of the system $s(t)$ and the target state $s_d$; i.e.,
$e(t)=s(t)-s_d$.
The Lyapunov function is thus defined as
\begin{align}
V(t)=e^T(t)Pe(t),
\end{align}
where $P\in\mathbb{R}^{3\times3}$ is a positive definite matrix. By using the dynamical equation of the coherent vector representation \eqref{vect}, the derivative of $V(t)$ with respect to time $t$ is calculated as
\begin{align}\label{dotV16}
\dot V(t)
&=2e^T(t)PAs(t)+2e^T(t)Pg\nonumber\\&+2u_1(t)e^T(t)PK_1s(t)+2u_2(t)e^T(t)PK_2s(t).
\end{align}
In order to ensure that $\dot V(t)<0$, we can eliminate the drift term $2e^T(t)PAs(t)+2e^T(t)Pg$ in the first row of \eqref{dotV16}  with the control function $u_{2}(t)$. Then, one has the following control mode
\begin{equation}\label{controller1}
M_1:\left\{
 \begin{aligned}
&u_1(t)=-\xi \mathrm{sign}[e^T(t)PK_1s(t)], \\
&u_{2}(t)=\frac{s^T(t)A^TPe(t)+e^T(t) Pg}{-e^T(t) PK_2 s(t)},
 \end{aligned}
 \right.
  \end{equation}
    where $\xi$ is a positive constant and $\mathrm{sign}(\cdot)$ is a sign function.
  Similarly, by employing the control function $u_{1}(t)$ to eliminate the drift term, a control mode can be formulated as
  \begin{equation}\label{controller2}
M_2:\left\{
 \begin{aligned}
&u_{1}(t)=\frac{s^T(t)A^TPe(t)+e^T(t) Pg}{-e^T(t) PK_1 s(t)} ,\\
&u_2(t)=-\xi \mathrm{sign}[e^T(t)PK_2s(t)].
 \end{aligned}
 \right.
  \end{equation}
Note that, although the above control laws are similar to those in Ref. \cite{19.11}, the target states considered in this paper are arbitrary, which challenges the results in Ref. \cite{19.11}. With the control mode \eqref{controller1} or \eqref{controller2}, we obtain
  \begin{align}\label{19}
\dot V(t)=-\delta \mathrm{sign}[e^T(t)PK_js(t)]e^T(t)PK_js(t)\leq0.
\end{align}
In Eq. \eqref{19}, it is less than or equal to zero, not just less than zero. This is because for the control mode $M_j$, there may exist solutions of $\dot V(t)=0$. According to the LaSalle invariance principle \cite{25,27.1}, the trajectory of the system state must converge to the invariant set $\hat S=\{s(t):\dot V(s(t))=0\}$. Hence, beside the target state which is generally in the invariant set, there would be other states in the invariant set, which can be an obstacle for state transition.

 In addition, due to the fractional form of the control laws, when the denominator approaches zero, the control function tends towards infinity, such that a singularity problem arises. The states that cause the control law \eqref{controller1} or \eqref{controller2} to become singular belong to a singular value set $ \hat T_j=\{s(t)|e(t)^TPK_{j}s(t)=0\}$, $j=1$ or $2$. From the definition of the singular value set, we find a target state belongs to this set since $e(t)=0$ when $s(t)=s_d$. However, the states in this set would not be the target state in general. Hence, the singular value sets and invariant sets form the primary obstacles in the state transfer, with entry into either of them making it challenging to reach the target state.
%
%
%

\subsection{Switching strategy}
To tackle the above problems, a switching control strategy is presented in this paper, whose main idea is to keep the system's state away from the singular value sets and invariant sets in the process of state transition.
 Given a terminal time $T_f>0$, we define the switching moments as
$\tau_m$ satisfying $0=\tau_0<\tau_1<\tau_2<\cdot\cdot\cdot<T_f$  where $m\in\mathbb{N}$.
 For $t\in[\tau_{m-1},\tau_m)$, the singular value set can be given as
\begin{align}
\hat T_{ \varkappa(m-1)}=\{s(t)|e^T(t)PK_{ \varkappa(m-1)}s(t)=0\},
\end{align}
where $ \varkappa(m-1)\in\{1,2\}$ labels the control mode during the interval $t\in[\tau_{m-1},\tau_m)$.
 $ K_{\varkappa(m-1)}$ represents the  control matrix over the interval $t\in[\tau_{m-1},\tau_m)$.
In order to avoid the singular value problem of control laws, reasonable thresholds $\kappa_1, \kappa_2>0$ are given to describe the distance to the singular value sets. Then, the switching moment caused by singular values is defined as
\begin{align}\label{tau1}
\tau^{sin}_m=\mathrm{inf}\{t\in[\tau_{m-1},T_f]:\Delta_{ \varkappa(m-1)}(t)-\kappa_{ \varkappa(m-1)}\leq0\}
\end{align}
with
\begin{align}
\Delta_{\varkappa(m-1)}(t)=|e^T(t)PK_{\varkappa(m-1)}s(t)|,
\end{align}
where $\kappa_{ \varkappa(m-1)}$ denotes the threshold for the distance from a state to the singular value set during $t\in[\tau_{m-1},T_f]$.
By switching controls at the moments \eqref{tau1}, we can effectively avoid the singular value sets.

Further, we consider the avoidance of invariant sets during the evolution. For $t\in[\tau_{m-1},\tau_m)$, the invariant set is given as
\begin{align}
\hat S_{ \varkappa(m-1)}=\{s(t)|\dot V(s(t), \varkappa(m-1))=0\}.
\end{align}
Since the system \eqref{vect} evolves slowly around an invariant set, it would take a long time for $\dot V_j(t)$ to be exactly equal to 0. To achieve a more rapid control response, two reasonable thresholds $\iota_1, \iota_2>0$ are given, and the moment of the switching caused by the invariant set can be defined as
\begin{align}\label{tau2}
\tau_m^{inv}=&\mathrm{inf}\{t\in[\tau_{m-1},T_f]:|\dot V(s(t), \varkappa(m-1))|\nonumber\\&-\iota_{ \varkappa(m-1)}\leq0\},
\end{align}
where $\iota_{ \varkappa(m-1)}$ represents the threshold for the distance from the state to the invariant set during the time interval $t\in[\tau_{m-1},T_f]$. Note that under different control modes, the invariant sets are different. Here we also use $\varkappa(m-1)\in\{1,2\}$ to label the control mode during the interval $t\in[\tau_{m-1},\tau_m)$.

Finally, combining \eqref{tau1} and \eqref{tau2}, we can write the switching time as
\begin{align}\label{tau3}
\tau_m&=\mathrm{min}\{\tau^{sin}_m,\tau_m^{inv}\}.
\end{align}
Then, the switching time sequence can be written as
\begin{align} \label{tau_vec}
\tau=\{\tau_1,\tau_2,...,\tau_N\},
\end{align}
where $N$ is the total amount of the switched times on time interval $[0,T_f]$.
 For the sake of analysis, the following assumption is given.

 \textbf{Assumption 1} Assume that the trajectory of the system will not pass through the intersections of singular value sets, nor through the intersections of the invariant sets.

 With Assumption 1, we avoid the situation where all control modes fail. Based on this assumption, we subsequently introduce a switching control algorithm.

\subsection{Switching control Algorithm}

It is clear that from the previous section, once the state evolves into the set $\hat T_{1}\cup \hat S_{1}$ or $\hat T_{2}\cup \hat S_{2}$ under control modes $M_1$ or $M_2$, the control mode switches to $M_{2}$ or $M_1$. With the strategy, we can summarize the switching control algorithm for an open qubit system as follows:
\\\textbf{Step 1.} Choose an arbitrary target state $s_d$ and initialize the state of the qubit system $s_0$. Set $m=0$, $j\in\{1,2\}$ and the appropriate thresholds  $\kappa_1, \kappa_2$ and $\iota_1, \iota_2$ for the singular value sets and invariant sets, respectively;
\\\textbf{Step 2.} A reasonable terminal condition $e^TPe\leq\vartheta$ with sufficiently small constant $\vartheta\in\mathbb R^+$ is given. If the termination condition is satisfied, stop the algorithm; if not, go to Step 3;
\\\textbf{Step 3.} If the invariant set condition or singular condition \eqref{tau3} is satisfied, take the current time $t$ as switching moment, set $m=m+1$ and $\tau_m=t$, one control mode is switched to the other one;
\\\textbf{Step 4.} Calculate the evolution of the state of the Markovian qubit system \eqref{vect} to the next state with one control mode $M_j$, $j\in\{1,2\}$. Return to Step 2.

With this algorithm, we can effectively prevent the system from entering invariant sets and singular value sets. In what follows, we analyze the convergence of the system's state to the target state under this algorithm.

\section{Main result}
 For arbitrary state transition in open qubit systems, the inability to achieve global asymptotic stability makes finite-time stability particularly significant. This is because it guarantees that, even in open environments, the system can reach the desired stable state within a finite time. This is particularly beneficial in suppressing decoherence in open quantum systems.
 In what follows, we provide a theorem on finite-time stability. Based on this, we propose sufficient conditions for finite-time contractive stability to achieve improved convergence.
%

\subsection{Finite-time stability}
Firstly, we shall introduce several concepts.

\textbf{Definition 1}\cite{25} Given constants $T_f, \varrho, \eta, c_1, c_2\in\mathbb{R}^+$ with $\varrho\in(0,T_f)$ and $\eta<c_1<c_2$, the system \eqref{vect} is said to be\\
(1) finite-time stable, if $|e(0)|<c_1$ implies that $|e(t)|<c_2$ for all $t\in[0,T_f]$.\\
(2) finite-time contractive stable,
if $|e(0)|<c_1$ implies that $|e(t)|<\eta$ for all $t\in[T_f-\varrho,T_f]$.

\textbf{Definition 2}\cite{25.1} For any $t_0<t_1<t_2$, let $N_\chi(t_1,t_2)$ be the switching times over the interval $[t_1,t_2)$. If $N_\chi(t_1,t_2)\leq N_0+(t_2-t_1)/\zeta$ holds for $N_0\leq1$, $\zeta>0$, then $\zeta$ and $N_0$ are called the average dwell time and the chatter bound, respectively.

It is worth mentioning that, from Definition 1, we can observe that the conditions for FTCS are much stronger than FTS.
Additionally, Definition 2 prevents frequent switching within a short period.
We take into account the scenario where $N_0=0$, which implies that no switching occurs when the time interval is shorter than $\zeta$.
Based on Definition 1 and Definition 2, we then have the following finite-time stability theorem.

\textbf{Theorem 1} If there exist matrices $P>0$, $W>0$ and constants $\alpha>0$, $c_1\geq|e(0)|$, $c_2>0$, such that
\begin{align}\label{th1}
T_f\leq-\frac{1}{\alpha} \mathrm{ln}\frac{\lambda_1c_1+d\lambda_3}{c_2\lambda_2},
\end{align}
where $\lambda_1=\lambda_{\mathrm{max}}(P)$, $\lambda_2=\lambda_{\mathrm{min}}(P)$, $\lambda_3=\lambda_{\mathrm{min}}(W)$, and the average dwell time $\zeta$ satisfies
\begin{align}\label{th2}
\zeta>\frac{T_f\mathrm{ln}\mu}{\mathrm{ln}(\lambda_2c_2)-\mathrm{ln}(\lambda_1c_1+d\lambda_3)-\alpha T_f},
\end{align}
where $\mu=\lambda_1/\lambda_2$, $d=g^TgT_f$, then the system \eqref{vect} is FTS with the control laws \eqref{controller1} and \eqref{controller2} under the strategy \eqref{tau3}. That is, the system state $s(t)$ converges to a neighborhood of target state $s_d$ with a distance less than $\sqrt c_2$ in a finite time $T_f$.

\textbf{Proof.}
To investigate the convergence of the system at the terminal time $T_f$, our initial focus is on analyzing the convergence behavior of the system between two switching moments. Since $W>0$ and $\alpha>0$, with the control laws \eqref{controller1} and \eqref{controller2}, it can be deduced that
\begin{align}\label{ghj}
\dot V_{ \varkappa(m)}(t)&=-\delta \mathrm{sign}(e^TPK_{ \varkappa(m)}s(t))e^TPK_{ \varkappa(m)}s(t)\nonumber\\
&\leq\alpha V_{ \varkappa(m)}(t)+g^T W g
\end{align}
for $t\in[\tau_m,\tau_{m+1})$.
According to \eqref{ghj}, the Lyapunov function satisfies
\begin{align}\label{Vtau11}
V_{ \varkappa(m)}(t)&<e^{\alpha(t-\tau_m)}V_{ \varkappa(m)}(e(t))\nonumber\\&+e^{\alpha t}\int^t_{\tau_m}e^{-\alpha w}g^TWg\mathrm{d}w \nonumber\\
&\leq e^{\alpha(t-\tau_m)}V_{ \varkappa(m)}(e(\tau_m))\nonumber\\&+e^{\alpha t}\int^t_{\tau_m}e^{-\alpha w}g^TWg\mathrm{d}w.
\end{align}
In view of $\lambda_1=\lambda_{\mathrm{max}}(P)$ and $\lambda_2=\lambda_{\mathrm{min}}(P)$, it is straightforward to obtain $e^T(t)Pe(t)\leq \lambda_1e^T(t)e(t)$ and $e^T(t)Pe(t)\geq \lambda_2e^T(t)e(t)$.
It then follows from the fact $\mu=\lambda_1/\lambda_2\geq1$  that
\begin{align}
e^T(t)Pe(t)\leq \mu e^T(t)Pe(t).
\end{align}
Next, we analyze the properties of the Lyapunov function at the switching moment $t=\tau_m$. Since the trajectory of $e(t)$ is continuous; i.e., $e(\tau_m)=e(\tau_m^-)$ with $e(\tau_m^-)=\mathrm{lim}_{\varpi\rightarrow0-}e(\tau_m+\varpi)$, it can be deduced that
\begin{align}
V_{ \varkappa(m)}(e(\tau_m))<\mu V_{ \varkappa(m-1)}(e(\tau_m^-)).
\end{align}
According to \eqref{Vtau11}, we have
\begin{align}
V_{ \varkappa(m+1)}(t)
&<e^{\alpha(t-\tau_{m+1})}V_{ \varkappa(m+1)}(e(\tau_{m+1}))\nonumber\\
&+e^{\alpha t}\int^t_{\tau_{m+1}}e^{-\alpha w}g^TWg\mathrm{d}w\nonumber\\
&\leq\mu e^{\alpha(t-\tau_{m+1})}V_{ \varkappa(m)}(e(\tau_{m}))\nonumber\\
&+e^{\alpha t}\int^t_{\tau_{m+1}}e^{-\alpha w}g^TWg\mathrm{d}w.\end{align}
For $t\in[0,T_f]$, we denote $N_\chi(0,t)$ as the total switching times in $[0,t)$ and obtain
\begin{align}
N_\chi(0,t)\leq N,
\end{align}
where $N$ is defined in \eqref{tau_vec}.
Next, we establish the relationship between the Lyapunov function and its initial values.
By iterating over the different time intervals, we have
\begin{align}\label{26}
V(t)&<e^{\alpha t}\mu^N V_{ \varkappa(0)}(0)+\mu^{N-1}e^{\alpha t}\int^{\tau_1}_0 e^{-\alpha w}g^TWg\mathrm{d}w\nonumber\\
&+ \mu^{N-2}e^{\alpha t}\int^{\tau_2}_{\tau_1}e^{-\alpha w}g^TWg\mathrm{d}w
+...\nonumber\\&+ e^{\alpha t}\int^{t}_{\tau_N}e^{-\alpha w}g^TWg\mathrm{d}w\nonumber\\
&=e^{\alpha t}\mu^N V_{ \varkappa(0)}(0)\nonumber\\&+ e^{\alpha t}\int^t_0e^{-\alpha w}\mu^{N_\chi(w,t)}g^TWg\mathrm{d}w.
\end{align}
According to Definition 2, we have $N\leq\frac{T_f}{\zeta}$. One can deduce that
\begin{align}
V(t)< e^{\alpha T_f}\mu^{\frac{T_f}{\zeta}}(V_{ \varkappa(0)}(0)+d\lambda_3).
\end{align}
From $\lambda_1=\lambda_{\mathrm{max}}(P)$, $\lambda_2=\lambda_{\mathrm{min}}(P)$,
 we obtain
\begin{align}\label{39}
e^T(t) e(t)\leq\frac{1}{\lambda_2}V(t)<\frac{1}{\lambda_2}\mu^{\frac{T_f}{\zeta}}e^{\alpha T_f}(\lambda_1c_1+d\lambda_3).
\end{align}
Noting that $\mu=\lambda_1/\lambda_2\geq 1$, we consider the following two distinct cases.
In the case of $\mu=1$, from \eqref{th1},
we finally get
\begin{align}
e^T(t)e(t)<c_2.
\end{align}
In the case of $\mu>1$, from \eqref{th2},  we obtain
\begin{align}
\frac{T_f}{\zeta}\leq\frac{\mathrm{ln}(\lambda_2c_2)-\mathrm{ln}(\lambda_1c_1+d\lambda_3)-\alpha T_f}{\mathrm{ln} \mu}.
\end{align}
Therefore, one can deduce that
\begin{align} \label{mu40}
\mu^\frac{T_f}{\zeta}&\leq
\mu^{\frac{\mathrm{ln}(\lambda_2c_2)-\mathrm{ln}(\lambda_1c_1+d\lambda_3)-\alpha T_f}{\mathrm{ln} \mu}}\\\nonumber
&=e^{\frac{\mathrm{ln}(\frac{\lambda_2c_2}{\lambda_1c_1+d\lambda_3})-\alpha T_f}{\mathrm{ln}\mu}\mathrm{ln}\mu}\\\nonumber
&=\frac{\lambda_2c_2}{ \lambda_1c_1+d\lambda_3}e^{-\alpha T_f}.
\end{align}
Combining \eqref{mu40} with \eqref{39}, we conclude that
\begin{align}
e^T(t)e(t)<c_2.
\end{align}
The proof is complete.  $\blacksquare$

\textbf{Remark 1}
The proof of Theorem 1 reveals that it is not necessary for $V(t)$ to be negative definite or semi-negative definite, which is unlike proving asymptotic stability. This distinction highlights the difference between finite-time stability and asymptotic stability.
In addition, Theorem 1 presents the conditions for the system to achieve FTS, where the bounds of the terminal state may not necessarily be smaller than the initial error. This result is different from that in \cite{19.2}, which can achieve global asymptotic stability as the target state is one of the eigenstates of the free Hamiltonian.

\subsection{Finite-time contractive stability}
In practice, it is often undesirable for the error range of the final state to exceed that of the initial state. To achieve a more precise terminal state, modified control laws can be designed as
\begin{equation}\label{controllerj1}
\hat M_1:\left\{
 \begin{aligned}
&u_1(t)=\frac{s^T(t)A^TPe(t)+e^T(t) Pg+\Upsilon(t)}{-e(t)^T PK_1 s(t)} \\
&u_{2}(t)=-\xi \mathrm{sign}(e^T(t)PK_{2}s(t))
 \end{aligned}
 \right.,
  \end{equation}
  \begin{equation}\label{controllerj2}
\hat M_2:\left\{
 \begin{aligned}
&u_1(t)=-\xi \mathrm{sign}(e^T(t)PK_{1}s(t))\\
&u_{2}(t)=
\frac{s^T(t)A^TPe(t)+e^T(t) Pg+\Upsilon(t)}{-e(t)^T PK_2 s(t)},
 \end{aligned}
 \right.
  \end{equation}
  where $\Upsilon(t)=-\Gamma(t)V(t)+\hat \theta $, $\xi>0$. $\Gamma(t)$ is a time-dependent function, $\hat \theta\in\mathbb{R}^+$. Based on this, a finite-time contractive stability theorem is given as follows.

\textbf{Theorem 2} Consider real functions $\Lambda(t)=\int \Gamma(t) \mathrm{d}t$, $\Gamma(t)<0$ and constants $\eta, b_1, \hat\theta\in\mathbb{R}^+$, $\varrho\in[0,T_f]$, $\eta<b_1$, $\alpha_1< \alpha_2\in\mathcal{K}$. If  we have
\begin{align}\label{con1}
\alpha_1(|e|)< V(|e|)<\alpha_2(|e|)
\end{align}
and
\begin{align}\label{con3}
e^{\Lambda(t)-\Lambda(\varrho)}\alpha_2(b_1)+\hat\theta e^{\Lambda(t)}\int^t_\varrho e^{-\Lambda(w)}\mathrm{d}w-\alpha_1(\eta)\leq0
\end{align}
with $|e(0)|\leq b_1$ for $\forall t\in[\varrho,T_f]$,
 the system \eqref{vect} is FTCS with the control laws \eqref{controllerj1} and \eqref{controllerj2} under the strategy \eqref{tau3}.

\textbf{Proof.} We first analyze the convergence of the system between two consecutive switching instants.
According to \eqref{controllerj1} and \eqref{controllerj2}, for
$t\in[\tau_m,\tau_{m+1})$, we have
\begin{align}
\dot V(t)&= \Gamma(t) V(t)-\hat\theta-\delta \mathrm{sign}(e^TPK_{ \varkappa(m)}s)e^TPK_{ \varkappa(m)}s.
\end{align}
Since $-\delta \mathrm{sign}(e^TPK_{ \varkappa(m)}s)e^TPK_{ \varkappa(m)}s\leq0$, we then have
\begin{align}\label{con2}
\dot V(t)\leq \Gamma(t) V(t)-\hat\theta.
\end{align}
Now, we consider the inequality conditions satisfied by the Lyapunov function between two switching moments. By solving the differential equation mentioned above, one can deduce that
\begin{align}\label{Vtau}
V(t)\leq e^{\int^t_{\tau_m}\Gamma(w)\mathrm{d}w}V(e(\tau_m))-\hat\theta e^{\Lambda(t)}\int^t_{\tau_m}e^{-\Lambda(w)}\mathrm{d}w.
\end{align}
Next, we establish the relationship between the Lyapunov function and $\alpha_1(\eta)$. In order to obtain convergence within the time period $t\in[\varrho,T_f]$, by iteratively applying \eqref{Vtau}, we have
\begin{align}
V(t)\leq e^{\Lambda(t)-\Lambda(\varrho)}V(e(\varrho))
-\hat\theta e^{\Lambda(t)}\int^{t}_{\varrho}e^{-\Lambda(w)}\mathrm{d}w.
\end{align}
It follows from \eqref{con3} that
\begin{align}
e^{\Lambda(t)-\Lambda(\varrho)}\alpha_2(b_1)
-\hat\theta e^{\Lambda(t)}\int^{t}_{\varrho}e^{-\Lambda(w)}\mathrm{d}w
&\leq \alpha_1(\eta).
\end{align}
 Noting that $V(e(\varrho))\leq V(e(0))<\alpha_2(b_1)$,
we finally have $\alpha_1(e(t))<V(e(t))< \alpha_1(\eta)$ with $\eta<b_1$ for $t\in[\varrho,T_f]$. Hence, the proof is complete.  $\blacksquare$

\textbf{Remark 2}
Theorem 2 establishes a more restrictive constraint compared to Theorem 1. The Bloch sphere can be viewed as a boundary accepted in Theorem 1, but this boundary is too loose for arbitrary state transition. In Theorem 2, a notable phenomenon emerges: the boundary $\eta$ and function $\alpha_1(\eta)$ exhibit a positive correlation. When the function $\Lambda(t)$ is characterized as a decreasing function, an intriguing consequence emerges. As the value of $e^{\Lambda(t)-\Lambda(\varrho)}$ decreases, the boundary of the system's state correspondingly contracts, which means that the boundary $\eta$ is much smaller than the initial error $b_1$.


\textbf{Remark 3}
Note that when the state of system \eqref{drho} is very close to the target state, both switching conditions in \eqref{tau1} and \eqref{tau2} can be satisfied. This may lead to the failure of the switching control strategy.
To address this issue, and in conjunction with the switching strategy in \eqref{tau3}, here we define shrink threshold functions
$\vartheta_i(V)$ and $\varsigma_i(V)$ to substitute the fixed thresholds $\kappa_i$ and $\iota_i$, respectively, where $\vartheta_i(\cdot)>0$ and $\varsigma_i(\cdot)>0$ are strictly increasing functions. Similar to \eqref{tau3}, we now give a switching control mechanism with shrink thresholds as follows
\begin{align}\label{tau4}
\tau_m&=\mathrm{inf}\{t\in[\tau_{m-1},T_f]:\mathrm{min}\{\Delta_{ \varkappa(m-1)}-\vartheta_{ \varkappa(m-1)}(V),\nonumber\\
&|\dot V(s(t), \varkappa(m-1))|-\varsigma_{ \varkappa(m-1)}(V)\}\leq0\}.
\end{align}
The equation above reveals that as the Lyapunov function $V$ continuously decreases, the corresponding threshold for  switching also decreases. This prolongs the duration of the control action.
Compared with the original switching strategy \eqref{tau3} which maintains fixed thresholds as the state approaches to the target state, the switching strategy \eqref{tau4} with shrink thresholds can achieve better performance.

\section{Simulation}
In this section, we verify our control performance in a single qubit system with several types of decoherence \cite{28}. The Hamiltonian of the single qubit is given as
$H_0=\frac{1}{2}\omega_0 \sigma_3$ with an angular frequency $\omega_0$. The control Hamiltonians are
$H_1=\sigma_1$,
$H_2=\sigma_2$, and the control matrices in the system \eqref{vect} are expressed as
 \begin{equation} \label{k1}
       K_1=\left[\begin{array}{ccc}0 & 0& 0\\0&0 &-2 \\0&2&0\\\end{array}\right],
        \end{equation}
       \begin{equation} \label{k2}
       K_2=\left[\begin{array}{ccc}0 & 0& 2\\0&0 &0 \\-2&0&0\\\end{array}\right].
        \end{equation}
        The initial state and the target state are given as
$
\rho_0=\left[\begin{array}{cc}0.8&0.4i  \\-0.4i &0.2\\\end{array}\right]
$ and $
\rho_d=\left[\begin{array}{cc}0.1&-0.3  \\-0.3 &0.9\\\end{array}\right].
$
It is worth noting that the target state, in this case, is not an eigenstate of the system's free Hamiltonian, $[H_0,\rho_d]\not=0$, nor that of the Lindblad operators, $[L,\rho_d]\not=0$, so that existing approaches for state transition in open qubit systems would fail in this example.
In what follows, we show the effectiveness of our switching control with different types of decoherence.
\subsection{The case of amplitude decoherence}
 Here, we consider a qubit system \eqref{drho} with one amplitude decoherence channel \cite{28.2} and let $\omega_0=10\mathrm{GHz}$ and $\gamma=0.1\mathrm{GHz}$. The Lindblad operator is given as
\begin{align}
L=\left[\begin{array}{cc}0&0  \\1 &0\\\end{array}\right].\nonumber
\end{align}
The matrix $A$ and the vector $g$ are given as
 \begin{align}
A=\left[\begin{array}{ccc}-0.05 & -10& 0\\ 10 &-0.05 &0 \\0&0&-0.1\\\end{array}\right],\nonumber
       \end{align}
        \begin{align}
g=\left[
         \begin{array}{c}
         0\\
         0\\
         -0.1\\
         \end{array}
       \right].\nonumber
       \end{align}
We set
$P=\mathrm{diag}[0.078,0.078,0.078]$ and $T_f=10\mathrm{a.u.}$. With the switching control laws \eqref{controllerj1} and \eqref{controllerj2}, we consider the control mode $\hat M_1$ as the initial control mode. The thresholds involving in \eqref{tau3} are given as $\kappa_1=0.0018$, $\kappa_2=0.00021$, $\iota_1=0.0047$ and $\iota_2=0.0001$. Moreover, in order to get a better control performance, we design a switching control strategy with a shrink threshold in \eqref{tau4}, where the shrink threshold functions are given as
$\vartheta_1(V)=0.0005V+0.0018$, $\vartheta_2(V)=0.00001V+0.000021$, $\varsigma_1(V)=0.00008V+0.0047$ and $\varsigma_2(V)=10^{-6}V$.

\begin{figure}[!t]
\centerline{\includegraphics[width=9cm]{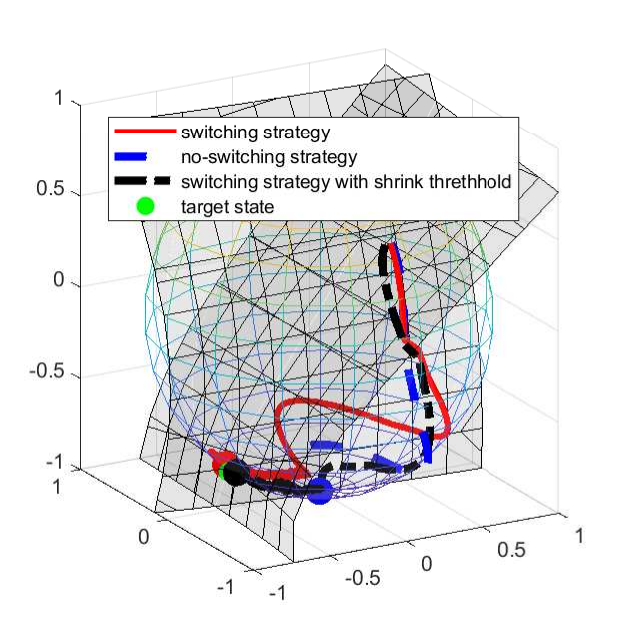}}
\caption{Comparison of trajectory evolution on Bloch sphere and two singular value sets}
\label{fig1}
\end{figure}

Fig. \ref{fig1} shows the evolution trajectories of the system's state on the Bloch sphere under switching control strategies. The blue dashed line is obtained without a switching control strategy, the red solid line represents the trajectory with a switching strategy \eqref{tau3}, and the black dashed line is the trajectory under the switching strategy \eqref{tau4} with a shrink threshold. Furthermore, the two intersecting gray surfaces represent the surfaces of two singular value sets of the system (which require switching when the trajectory crosses them), with the target state (green point) lying on the intersection curve.
In Fig. \ref{fig1}, it can be observed that without a switching strategy, the system ultimately converges to the invariant set of the system (blue point), making it unable to reach the neighborhood of the target state. On the contrary, with the switching control strategy, the terminal states (red point and black point) can cross the invariant set to reach the neighborhood of the target state. Additionally, Fig. \ref{fig1} reveals that the black point is closer to the green point, suggesting that switching strategy \eqref{tau4} is superior to switching strategy \eqref{tau3}.

\begin{figure}[!t]
\centerline{\includegraphics[width=10cm]{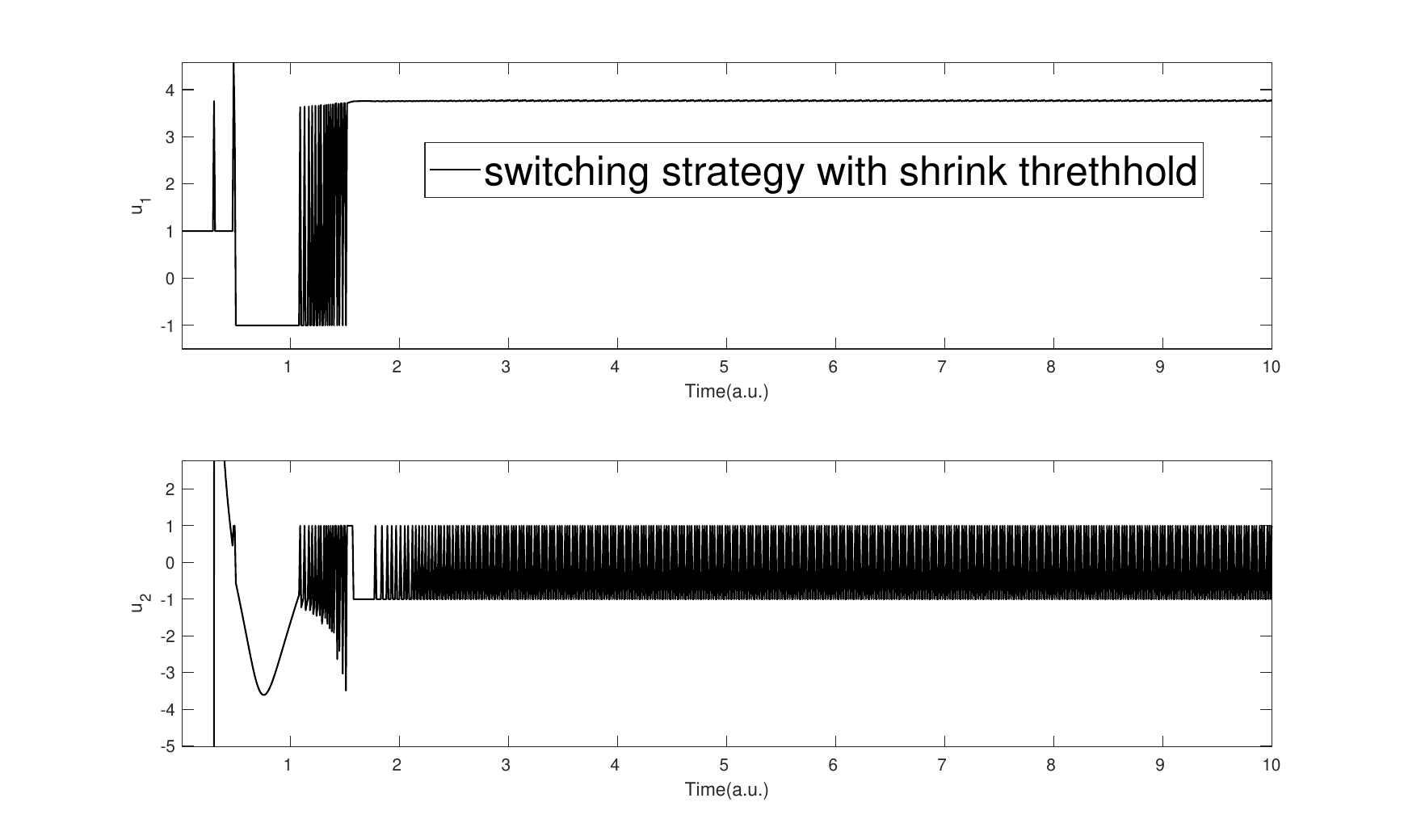}}
\caption{Control curves on two control channels}
\label{fig1.01}
\end{figure}

 In addition, control laws $u_1(t)$ and $u_2(t)$ under switcing control strategy with shrink thresholds are given in Fig. \ref{fig1.01}. We find that control laws in this paper are continuous and non-smooth. Furthermore, the curve of the control function $u_2(t)$ oscillates in the later stage, which is not caused by the switching of the control, but because of the sign function in \eqref{controllerj1} and \eqref{controllerj2}, similar to the high-frequency pulse function in dynamical decoupling \cite{28.1}.



At the terminal time $T_f=10 \mathrm{a.u.}$, under the switching control strategy with the shrink threshold, the density matrix of system is given as
\begin{align}
\rho(T_f)=\left[\begin{array}{cc}0.10&-0.29+0.03i  \\-0.29-0.03i&0.90\\\end{array}\right]\nonumber
\end{align}
and its fidelity is 99.4\%. According to Theorem 2, we set $\Lambda(t)=-kt$, $k=1.5$, $\hat\theta=-0.1g^Tg$. The eigenvalue $\lambda_2=0.078$ can be obtained from the matrix $P$. We choose $\alpha_2(e)=0.08e^2$ and $\varrho=7\mathrm{a.u.}$.
 It can be deduced that $\eta=\sqrt{\frac{1}{\lambda_2} [e^{\Lambda(t)-\Lambda(\varrho)}\alpha_2(b_1)+\hat\theta e^{\Lambda(t)}\int^t_\varrho e^{-\Lambda(s)}ds]}\approx 0.06$. According to Theorem 2, $\eta$ is an upper bound on the error distance of the state within the interval $[7, 10]$. From \eqref{con1}, $\alpha_2(\eta)$ is the upper bound of Lyapunov function within the interval $[7, 10]$.

\begin{figure}[!t]
\centerline{\includegraphics[width=9cm]{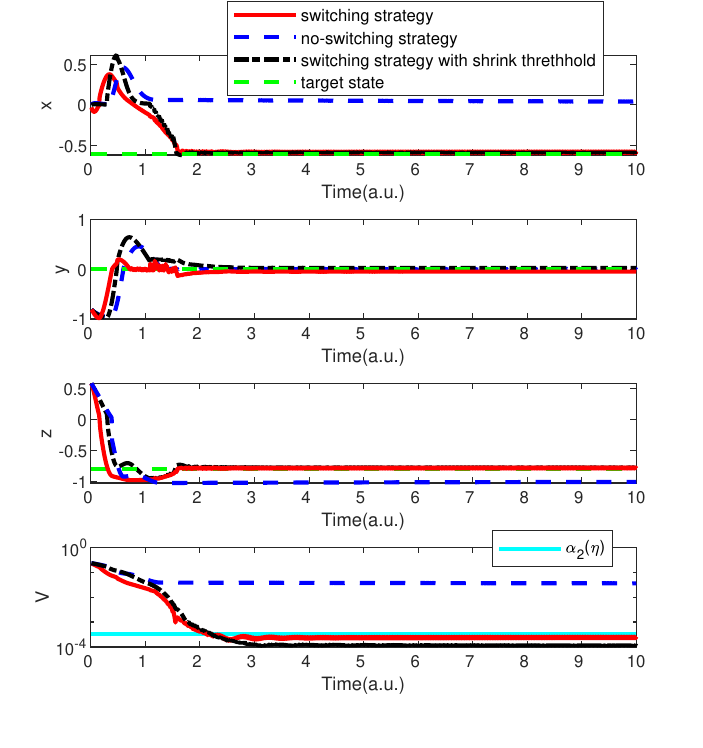}}
\caption{Comparison of different evolutionary trajectories and Lyapunov functions}
\label{fig1.1}
\end{figure}

  In Fig. \ref{fig1.1}, we compare the trajectory of states under different control strategies. It is clear that the non-switching control strategy maintains a significant distance in the $x$ direction, thereby failing to enter the neighborhood of the target state. In contrast, the switching control strategy effectively addresses this issue. Notably, the Lyapunov curve under the switching control strategy (red solid line) is less than the theoretical upper bound $\alpha_2(\eta)$ within the interval $[7, 10]$, as defined in Theorem 2.
  Additionally, the Lyapunov function under the switching control strategy with a shrink threshold (black dashed line) is lower than that under the fixed threshold switching control strategy (red solid line). This indicates that the shrink threshold control strategy proposed in \eqref{tau4} can enhance convergence.


\subsection{The case of dephasing decoherence}
\begin{figure}[!t]
\centerline{\includegraphics[width=8cm]{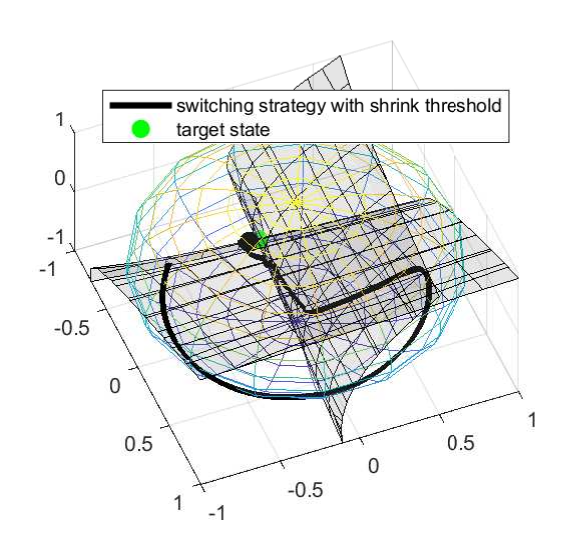}}
\caption{The state trajectory in the case of dephasing decoherence}
\label{fig7}
\end{figure}

\begin{figure}[!t]
\centerline{\includegraphics[width=9cm]{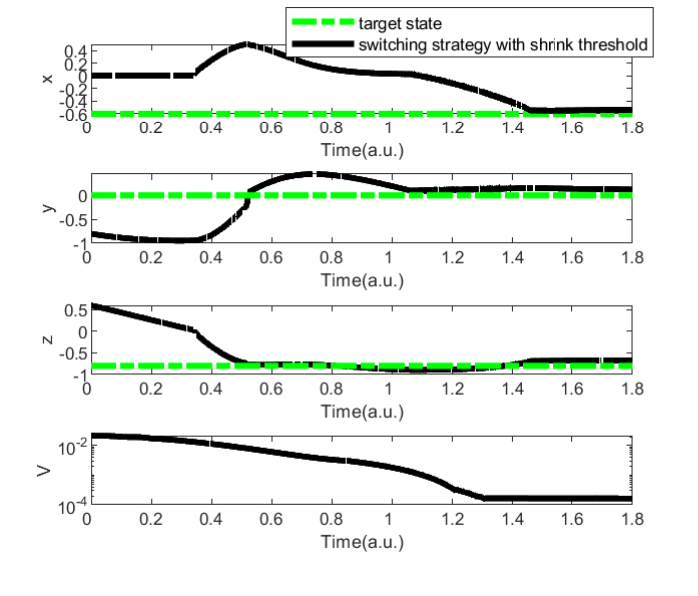}}
\caption{State trajectory and Lyapunov function  in the case of dephasing  decoherence}
\label{fig7.1}
\end{figure}

When taking into account dephasing decoherence \cite{28.2}, the Lindblad operator can be expressed as $
L=\left[\begin{array}{cc}1&0  \\0 &-1\\\end{array}\right]$.
Here, we select $\omega_0=10\mathrm{GHz}$ and $\gamma=0.1\mathrm{GHz}$.
 The  system matrix is obtained as
 \begin{align}
A=\left[\begin{array}{ccc}-0.2 & -10& 0\\ 10 &-0.2 &0 \\0&0&0\\\end{array}\right].
       \end{align}
       The control matrices are the same as \eqref{k1} and \eqref{k2}. The vector $g$ is a zero vector.
       Likewise, we consider the switching control strategy with shrink thresholds, where the shrink threshold functions are given as
$\vartheta_1(V)=0.3V$, $\vartheta_2(V)=0.4V+0.00035$, $\varsigma_1(V)=1.2V+0.0002$ and $\varsigma_2(V)=10^{-6}V$ in \eqref{tau4}. Take the control mode $\hat M_1$ as the initial control mode. Set $P=\mathrm{diag}[0.078,0.078,0.078]$ and $T_f=1.8\mathrm{a.u.}$. The simulation results are shown in Figs. \ref{fig7} and Figs. \ref{fig7.1}.


Fig. \ref{fig7} presents the evolutionary trajectory of the system's state with the shrink threshold switching control strategy. The grey surfaces represent invariant sets and singular value sets. It is evident that when the system passes through these sets, it is neither attracted by them nor disrupted by any singular values, maintaining a continuous trajectory. Ultimately, the system's terminal state is able to reach a neighborhood of the target state.
In Fig. \ref{fig7.1}, we observe that the system state converges to a neighborhood of the target state in all directions. Furthermore, in Fig. \ref{fig7.1}, we find that the Lyapunov function converges to an order of magnitude of $10^{-4}$ at terminal time. This demonstrates the effectiveness of the methods presented in this paper in the dephasing decoherence.

\subsection{The case of polarization decoherence}
Now, we consider the switching control strategy in polarization decoherence case \cite{28.3}. Three damping channels are considered here and the damping rates are given as $\gamma_1=\gamma_2=\gamma_3=0.01\mathrm{GHz}$. Three different Lindblad terms are given as
$
L_1=\sigma_3$, $
L_2=\sigma_2$ and $
L_3=\sigma_1$
Here, we set $\omega_0=10\mathrm{GHz}$ and the system matrix can be expressed as
 \begin{align}
A=\left[\begin{array}{ccc}-0.04 & -10 & 0\\ 10 &-0.04 &0 \\ 0&0& -0.04\\\end{array}\right].
       \end{align}
The vector $g$ is a zero vector. Set $T_f=2\mathrm{a.u.}$ and take the control mode $\hat M_1$ as the initial control mode.
Based on Remark 3, we have designed shrink thresholds for the switching control strategy.
The shrink threshold functions in \eqref{tau4} are given as
$\vartheta_1(V)=0.01V+0.001$, $\vartheta_2(V)=0.01V+0.002$, $\varsigma_1(V)=0.001V$ and $\varsigma_2(V)=10^{-6}V$.
It is observed that as the state evolves towards the target state, the error $e(t)$ continuously diminishes, leading to a corresponding reduction in the Lyapunov function $V(t)$. This, in turn, results in a contraction of the threshold. This effectively prolongs the duration of the switching control strategy, allowing us to obtain a better control performance.

\begin{figure}[!t]
\centerline{\includegraphics[width=8cm]{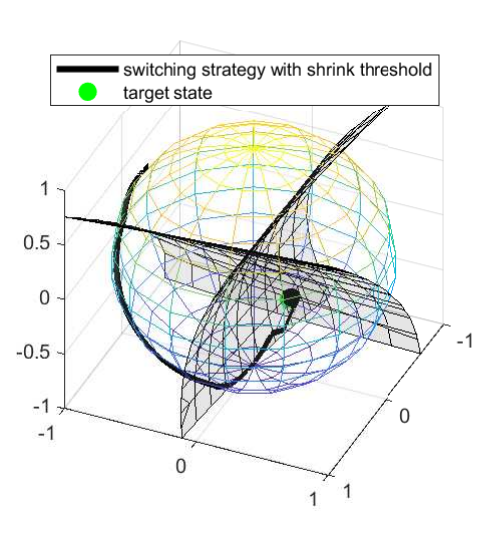}}
\caption{The state trajectory on the Bloch sphere in the case of polarization decoherence}
\label{fig11}
\end{figure}
\begin{figure}[!t]
\centerline{\includegraphics[width=9cm]{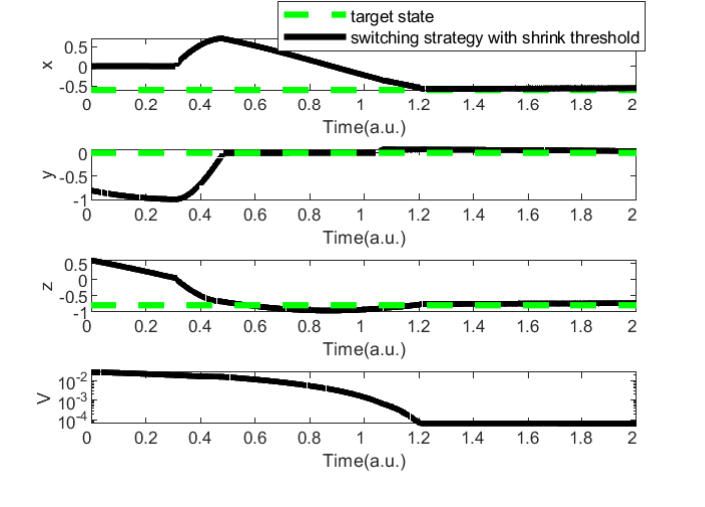}}
\caption{The state trajectory and Lyapunov function in the case of polarization decoherence}
\label{fig11.1}
\end{figure}

Fig. \ref{fig11} and Fig. \ref{fig11.1} depict the convergence of the target state under the shrink threshold switching control strategy. As illustrated in Fig. \ref{fig11}, under the switching control strategy, the system's state trajectory can pass through invariant sets and  singular value sets to reach the neighborhood of the target state. Fig. \ref{fig11.1} shows the convergence of the system state in three directions. Ultimately, the Lyapunov function's curve reveals that the designed control laws can stabilize the system state within the neighborhood of the target state. Therefore, we can conclude that, similar to the cases of amplitude and dephasing decoherence, the strategy proposed in this paper remains effective in the polarization decoherence.

\section{Conclusion}
This paper presents a switching control strategy for arbitrary state transitions in open qubit systems. The switching control strategy prevents the state of open qubit systems from invariant and singular value sets. Furthermore, we have demonstrated that the proposed switching control strategy can enable the system to achieve FTS. Additionally, to enhance system convergence, we have proposed an improved controller and outlined the conditions for achieving finite-time contractive stability. Finally, the effectiveness of this method is validated through three examples. It is worth noting that our methods can be applied to quantum systems in three different decoherence scenarios, significantly expanding its applicability. Future research directions involve extending qubit systems to $n$-level systems with measurement feedback.
\end{document}